\documentclass[a4paper,12pt]{article}
\usepackage[]{epsfig}
\usepackage[]{rotating}
\usepackage[]{cite}
\usepackage{subfigure}
\usepackage{booktabs}
\newcommand{\inmath}[1] {\ifmmode#1\else$#1$\fi}
\newcommand{\definmath}[2] {\def#1{\ifmmode#2\else$#2$\fi}}
\definmath{\Pq}      {\mathrm{q}}
\definmath{\Paq}  {\overline{\mathrm{q}}}
\definmath{\Pb}      {\mathrm{b}}
\definmath{\Pc}      {\mathrm{c}}
\definmath{\Pac}  {\overline{\mathrm{c}}}
\definmath{\Pu}      {\mathrm{u}}
\definmath{\Pau}  {\overline{\mathrm{u}}}
\definmath{\Pd}      {\mathrm{d}}
\definmath{\Pad}  {\overline{\mathrm{d}}}
\definmath{\Ps}      {\mathrm{s}}
\definmath{\Pas}  {\overline{\mathrm{s}}}
\definmath{\dEdx} {{\mathrm d}E/{\mathrm d}x}
\definmath{\4f}      {\mathrm{4f}}
\definmath{\Pab}  {\overline{\mathrm{b}}}

\newcommand{\bbbar}  {\Pb\Pab}
\newcommand{\uubar}  {\Pu\Pau}
\newcommand{\ddbar}  {\Pd\Pad}
\newcommand{\ssbar}  {\Ps\Pas}
\newcommand{\massof}[1] {m_{\smash{#1}\mathstrut}}
\newcommand{\ccbar}  {\Pc\Pac}
\newcommand{\Rb}  {R_{\mathrm{b}}}
\newcommand{\Ruds}  {R_{\mathrm{uds}}}
\newcommand{\Rc}  {R_{\mathrm{c}}}
\newcommand{\PDz} {\rm D^0}
\newcommand{\PDp} {\rm D^+}
\newcommand{\PDsp} {\rm D_s^+}
\definmath{\PDstpm} {{\mathrm{D}^{*\pm}}}     
\newcommand{\qqbar}  {\Pq\Paq}

\newcommand{\mPZ} {\massof{\mathrm{Z}}}
\newcommand{\mtop}   {\massof{\mathrm{top}}}
\newcommand{\PZz}  {\rm Z}
\newcommand{\mHiggs} {\massof{\mathrm{Higgs}}}
\newcommand{\alphas} {\alpha_{\mathrm{s}}}
\newcommand{\gbb} {$\mathrm{g_{b\overline b}}$}
\newcommand{\gcc} {$\mathrm{g_{c\overline c}}$}
\newcommand{\gbbd} {\mathrm{g_{b\overline b}}}
\newcommand{\gccd} {\mathrm{g_{c\overline c}}}

\newcommand{\KORALW}{\mbox{K{\sc oral}W}}

\newcommand{\PYTHIA}{\mbox{P{\sc ythia}}}

\newcommand{\ZFITTER}{\mbox{Z{\sc fitter}}}
\newcommand{\PhysRev}   {Phys.~Rev.}

\newcommand{\NIM} {Nucl.~Instr.\ and Meth.}
\newcommand{\ZPhys}  {Z.~Phys.}

\newcommand{\CPC} {Comp. Phys. Comm.}
 
\newcommand{\epem}   {\rm e^+ e^-}

\newcommand{\wpwm}   {\rm W^+ W^-}

\def\etal{\mbox{{\it et al.}}}

\begin{document}

\pagenumbering{roman} \pagestyle{plain}
\begin{titlepage}
\pagenumbering{arabic}  \setcounter{page}{1}
\begin{center}{\large   EUROPEAN ORGANIZATION FOR NUCLEAR RESEARCH
}\end{center}\bigskip
\begin{flushright}
       CERN-PH-EP/2004-051   \\ 13 September 2004
\end{flushright}
\bigskip\bigskip\bigskip\bigskip\bigskip
\begin{center}{\huge\bf   Measurements of $\Rb$ in $\epem$ Collisions at 182 -- 209 GeV
}\end{center}\bigskip\bigskip
\begin{center}{\LARGE The OPAL Collaboration
}\end{center}\bigskip\bigskip
\bigskip\begin{center}{\large  Abstract}\end{center}
Measurements of $\Rb$, the ratio of the $\bbbar$ cross-section to the $\qqbar$ 
     cross-section in $\epem$ collisions, are presented. The data were 
     collected by the OPAL experiment at LEP at centre-of-mass energies 
     between 182 GeV and 209 GeV. Lepton, lifetime and event-shape 
     information is used to tag events containing b quarks with high
     efficiency. The data are compatible with the Standard Model expectation.
     The mean ratio of the eight measurements reported here 
     to the Standard Model prediction is 
      $1.055\pm0.031\pm0.037$, where the first error is 
     statistical and the second systematic.
\bigskip\bigskip\bigskip\bigskip
\bigskip\bigskip
\begin{center}{\large
(To be Submitted to Physics Letters B)
}\end{center}

\end{titlepage}

\begin{center}{\Large        The OPAL Collaboration
}\end{center}\bigskip
\begin{center}{
G.\thinspace Abbiendi$^{  2}$,
C.\thinspace Ainsley$^{  5}$,
P.F.\thinspace {\AA}kesson$^{  3,  y}$,
G.\thinspace Alexander$^{ 22}$,
J.\thinspace Allison$^{ 16}$,
P.\thinspace Amaral$^{  9}$, 
G.\thinspace Anagnostou$^{  1}$,
K.J.\thinspace Anderson$^{  9}$,
S.\thinspace Asai$^{ 23}$,
D.\thinspace Axen$^{ 27}$
% ...
}\end{center}\bigskip
\bigskip
$^{  1}$School of Physics and Astronomy, University of Birmingham,
Birmingham B15 2TT, UK
\newline
$^{  2}$Dipartimento di Fisica dell' Universit\`a di Bologna and INFN,
I-40126 Bologna, Italy
\newline
$^{  3}$Physikalisches Institut, Universit\"at Bonn,
D-53115 Bonn, Germany
\newline
% ...
$^{  5}$Cavendish Laboratory, Cambridge CB3 0HE, UK
\newline
% ...
$^{  9}$Enrico Fermi Institute and Department of Physics,
University of Chicago, Chicago IL 60637, USA
\newline
% ...
$^{ 16}$Department of Physics, Schuster Laboratory, The University,
Manchester M13 9PL, UK
\newline
% ...
$^{ 22}$Department of Physics and Astronomy, Tel Aviv University,
Tel Aviv 69978, Israel
\newline
$^{ 23}$International Centre for Elementary Particle Physics and
Department of Physics, University of Tokyo, Tokyo 113-0033, and
Kobe University, Kobe 657-8501, Japan
\newline
% ...
$^{ 27}$University of British Columbia, Department of Physics,
Vancouver BC V6T 1Z1, Canada
\newline
% ...
$^{  y}$ now at CERN
\newline
% ...

\section{Introduction}
The cross-section for bottom-quark pair production in $\epem$
annihilation relative to the hadronic cross-section,
$$\Rb \equiv \frac{\sigma( \epem \to \gamma / \rm{Z}  \to \bbbar   )}
{\sigma( \epem \to  \gamma / \rm{Z} \to \qqbar)},$$ 
is a sensitive probe of the Standard Model \cite{bib:sm-rb}.
Measurements of $\Rb$ have been made at the $\PZz$ peak and at higher
energies \cite{bib:ew}.
At the $\PZz$ resonance, where fermion-pair production is dominated by $\PZz$
decays, measurements of $\Rb$ provide a precise determination of the ratio of the 
$\PZz\to\bbbar$
partial width to the hadronic width\footnote{$\Rb^0$ is the partial width ratio 
for $\PZz$ decays and not the cross-section ratio measured in this paper. 
It is smaller than the cross-section ratio at the peak of the $\PZz$ resonance 
by 0.0002.} $R^0_{\rm b} = \Gamma_{\rm \bbbar}/\Gamma_{\rm had}$.
This quantity is of particular interest because of its unique sensitivity 
to electroweak radiative corrections; while
sensitive to the top-quark mass, its dependence on other parameters, for 
example the Higgs boson mass and the strong coupling constant, is 
negligible \cite{bib:ew}.
Above the  $\PZz$ peak, the pure  $\PZz$ cross-section decreases and the
contributions of photon exchange and $\gamma$--Z interference become important.
Possible new physics at a high energy scale might manifest itself as a
deviation from the Standard Model prediction.

   In this paper, measurements of $\Rb$ at energies above the $\PZz$ resonance
are presented.  The data were taken by the OPAL detector at the LEP $\epem$
collider, at centre-of-mass energies, $\sqrt s$, ranging from 182 GeV to 209 GeV,
during the LEP2 programme.

   Above the $\PZz$ peak a significant fraction of the observed fermion-pair
events comes from radiative return to the $\PZz$ through initial-state photon
radiation. Only non-radiative events are considered here, according to
the definition used by OPAL in the analysis of fermion-pair production
at LEP2 \cite{bib:twofermions}:
\begin{itemize}
\item An effective centre-of-mass energy  $\sqrt{s'}>0.85\sqrt{s}$, where $s'$ is defined as the square of the mass of the $\gamma / \rm{Z} $ propagator.
\item The predicted contribution of interference between initial- and final-state photon 
radiation is removed. 
\end{itemize}

At high energies, additional background sources also arise, mainly from
$\wpwm$ and $\PZz\PZz$ decaying to four-fermion final states.  

  Because of the limited statistics due to the low cross-section at LEP2,
the double-tag technique as used at the $\PZz$ resonance 
\cite{bib:double-tag} is not optimal.
Instead the measurement reported here relies on a single-tag method, and
uses a sophisticated tagging algorithm based on lepton, lifetime and
event-shape information to identify $\bbbar$ events.  This algorithm is
more efficient and has a higher purity than the one used in our previous
measurement \cite{bib:pr297}, where $\Rb$ was measured up to 189 GeV. Hence the
183 GeV and 189 GeV results reported here supersede the previous values, while
the measurements at higher energy are reported here by OPAL for the first
time.

\section{The OPAL detector, data sample and simulation}
A detailed description of the OPAL detector can be found
elsewhere \cite{bib:OPALdetectoratLEPtwo}. For this analysis,
the most relevant parts of the
detector are the silicon micro-vertex detector, the
tracking chambers, the
electromagnetic and hadronic calorimeters, and the muon chambers.
The micro-vertex detector is essential for the reconstruction of secondary
vertices. The central
detector provides precise measurements of the momenta of charged particles
from the curvature of their trajectories
in a magnetic field of $0.435\;$T. In addition, it allows the identification
of charged particles through a combination of the measurement of the
specific energy loss $\dEdx$ and the momentum.
The electromagnetic calorimeter consists
of 11\,704 lead glass blocks, which completely cover
the azimuthal\footnote{The OPAL coordinate
system is defined as a right-handed Cartesian coordinate system, with the
$x$ axis pointing in the plane of the LEP collider toward the
centre of the ring, the $z$ axis in the direction of the
electron beam, and $\theta$ and $\phi$ defined as the usual spherical
polar coordinates.} range up to polar angles
of $|\cos \theta|<0.98$. Almost
the entire detector is surrounded with four layers of
muon chambers, after approximately one metre of iron
from the magnet return yoke, which is instrumented as a hadron calorimeter.
Luminosity is determined using small-angle Bhabha scattering in the forward calorimeter \cite{bib:twofermions}.
\subsection {Data sample}
The data used in this analysis were collected by the OPAL
detector at LEP during 1997-2000, at centre-of-mass energies
between 182 GeV and 209 GeV. They are grouped into eight energy
points. The centre-of-mass energies and integrated luminosities of the data samples
are summarized in Table~\ref{tab:R4f}.

\begin{table}[htb]
\def\pz{\phantom{0}}
 \begin{center} \begin{tabular}{c|c|c|c} \toprule
Nominal Energy & Energy Range (GeV) &$\langle\sqrt{s}\rangle$ (GeV) & Integrated Luminosity ($\rm{pb}^{-1}$)\\
\bottomrule
183 & 181.7 - 184.0	& 182.7 &\pz 56 \\
189 & 188.5 - 189.1	& 188.6 &185 \\
192 & 191.4 - 191.7	& 191.6 &\pz 29  \\
196 & 195.4 - 195.8	& 195.5 &\pz 77  \\
200 & 199.3 - 199.9	& 199.5 &\pz 78  \\
202 & 201.5 - 202.0	& 201.6 &\pz 36  \\
205 & 204.5 - 205.5	& 205.3 &\pz 74  \\
207 & 205.5 - 208.6	& 206.8 &137 \\
\bottomrule
 \end{tabular}
 \caption{The nominal energy, energy range, luminosity-weighted mean centre-of-mass energy and integrated luminosity of the data used in the analysis.}
\label{tab:R4f}
 \end{center}
\end{table}

\subsection{Simulation}
\label{simulation}

Monte Carlo (MC) simulation is used to estimate the selection efficiency for $\bbbar$ 
events and to evaluate backgrounds. 
MC samples 
are generated at the  nominal centre-of-mass energies given in Table~1.
The KK2F program \cite{bib:kk2f} is used to simulate $\epem\to\qqbar(\gamma)$ events.
Four-fermion background
events are simulated with the grc4f generator \cite{bib:grc4f} or with the 
\KORALW\ \cite{bib:koralw} program with the grc4f matrix elements. 
The hadronization 
is performed by the \PYTHIA\ 6.150 \cite{bib:pythia} string model, 
and 
heavy quark fragmentation is modelled according to the Peterson fragmentation 
scheme \cite{bib:peterson} with parameters tuned according 
to \cite{bib:frag-parameters}. 
All the MC samples are passed through a detailed simulation of the OPAL 
detector \cite{bib:gopal}.

The Standard Model cross-section predictions are calculated using $\ZFITTER$ 6.36 
\cite{bib:zfitter}. $\ZFITTER$ is used to calculate the Standard Model value of $\Rc$, 
the ratio of the 
cross-section for charm-quark pair production to the hadronic cross-section,
at which the resulting $\Rb$ is quoted, and to 
calculate $\Rb$ for comparison of the 
theoretical predictions with the measured values. 
$\ZFITTER$ is also used for estimating the effects of interference between 
initial- and final-state  photon radiation, as described in Section \ref{ifi}.\\

\section{Analysis procedure}
\label{analysis}
The measurement of $\Rb$ starts by selecting non-radiative hadronic
    events, applying fiducial cuts and rejecting four-fermion background events.
The number of selected events, $N_{\rm had}$, is corrected for background
    and for the efficiency of the selection cuts to yield a corrected number of 
non-radiative hadronic events, 
$N_{\qqbar}$. The background includes the residual
four-fermion
    background and feedthrough of events with lower effective centre-of-mass energy, $\sqrt{s'}$. The contribution from interference
    between initial- and final-state radiation is also removed. A b-tagging algorithm is 
    applied to the selected events, and the number of tagged events, $N_{\rm tag}$, is 
    similarly corrected for the same background contributions and interference 
to obtain $N_{\bbbar}$, which still contains contributions from $\ccbar$ 
and light-quark-pair events. 
$\Rb$ is then calculated using:
 $$\Rb={N_{\bbbar}/N_{\rm \qqbar} - \epsilon_{\rm c} \Rc -(1-\Rc)\epsilon_{\rm uds}\over 
\epsilon_{\rm b}-\epsilon_{\rm uds}},\eqno{(1)}$$ where 
$\epsilon_{\rm b,c,uds}$ are the efficiencies 
for non-radiative $\bbbar$, $\ccbar$ and light-quark-pair 
events to pass the selection criteria and the b-tagging algorithm.

In the following sections, we describe the selection of high quality non-radiative 
hadronic events,
the rejection of four fermion background (Section \ref{evsel}), and the tagging 
algorithm used to select events originating from $\epem\to\bbbar$ 
(Section \ref{btag}).
This is followed by the description of the residual background estimation 
(Section \ref{ff}). 
The correction for interference is discussed in Section \ref{ifi}, and finally the 
calculation of $\Rb$ is presented (Section \ref{results}). The systematic 
uncertainties are described in Section \ref{sys}, and
the final results and conclusion are given in Section \ref{conclusion}.


\subsection{Hadronic event selection}
\label{evsel}
Hadronic events, $\epem\to \gamma / \rm{Z} \to\qqbar$, are selected based on the 
number of reconstructed charged particle tracks and the energy deposited in the 
calorimeters \cite{bib:hsel}. Further requirements follow, 
demanding at least seven tracks and that 
the polar angle of the thrust axis, $\theta_{\rm thr}$,
satisfy $|\cos\theta_{\rm thr}|<0.9$; the thrust axis is calculated 
from all tracks 
and clusters and corrected for double-counting as in \cite{bib:172opal}. 
The effective centre-of-mass energy, $\sqrt{s'}$, of the $\epem$
collision is estimated as described in Ref. \cite{bib:twofermions}, and 
non-radiative hadronic events are selected if $\sqrt{s'/s}>0.85$.
For the centre-of-mass energies analysed in this paper, the production cross-section 
of $\wpwm$ events  is comparable 
to that of non-radiative $\qqbar$ events. The production cross-section of $\PZz\PZz$ events is about an 
order of magnitude smaller.  Using the same techniques as in 
\cite{bib:pr321}, events are rejected if they are identified as 
$\wpwm$ events.

The number of events at each energy selected by the above cuts ($N_{\rm had}$) is 
given in Table~\ref{tab:results}.
Typical efficiencies, including the $\approx 90\%$ efficiency of the $\cos\theta_{\rm thr}$ 
cut, are $77\%$ for $\bbbar$ and 80\% for lighter
(u, d, s and c) $\qqbar$ events. About 14\% of the $N_{\rm had}$ sample are expected to be 
$\bbbar$ events.
 The selection of non-radiative events is less efficient for $\bbbar$ final 
states than for 
other flavours because of the generally larger missing energy due to the neutrinos in 
semileptonic b and c
hadron decays.
The remaining four-fermion background is estimated to be about 6\% of the selected hadronic events.

\subsection{b-tagging}
\label{btag}
In  $\epem\to\gamma / \rm{Z}  \to\bbbar$ events, the quark and the anti-quark are 
typically boosted 
in opposite directions, and 
the subsequent hadronization is largely independent. In this analysis, each event is 
divided into two hemispheres 
defined by a plane that is orthogonal to the thrust direction. 

We use a hemisphere-based b-tagger designed for LEP2 Higgs searches  \cite{bib:pr367}, 
where the tagging of $\bbbar$ events is based on three nearly independent properties of 
the b-hadron and its decay products:
 a lepton from a semileptonic b-hadron decay, the long lifetime, and kinematic 
differences between b-hadron decays and fragmentation in $\uubar,\ddbar,\ssbar$ events,
due to the hard 
fragmentation 
of the b-hadron and the high multiplicity of its decay products. 

For the lepton tag, semileptonic b-hadron decays are identified from their resultant 
electron or muon and the lepton momentum
is used as the b-tag variable.
Electrons are identified using an artificial neural network 
(ANN) \cite{bib:double-tag} while muons are identified using
information from the muon chambers in association with the tracking chambers, as
 in~\cite{muon}.
Electrons from photon conversions are rejected using an additional ANN described in
\cite{bib:pr146}. 

For the lifetime tag, secondary vertices are reconstructed. Variables
based on the vertex significance (the distance between the reconstructed primary and 
secondary vertices divided by the corresponding uncertainty)
and the impact parameters of the tracks associated 
with the vertex are combined using an ANN scheme as in \cite{bib:pr367}.

For the kinematic tag,
three variables are combined using another ANN:
the number of detected particles in the central part 
of the hemisphere, the angle between the hemisphere axis and its boosted sphericity 
axis,
and the C parameter \cite{bib:Cpar}, the latter two being calculated in the rest frame
of all the particles associated with the hemisphere.

\begin{figure}
\centering
{ 
    \includegraphics[width=6.8 in]{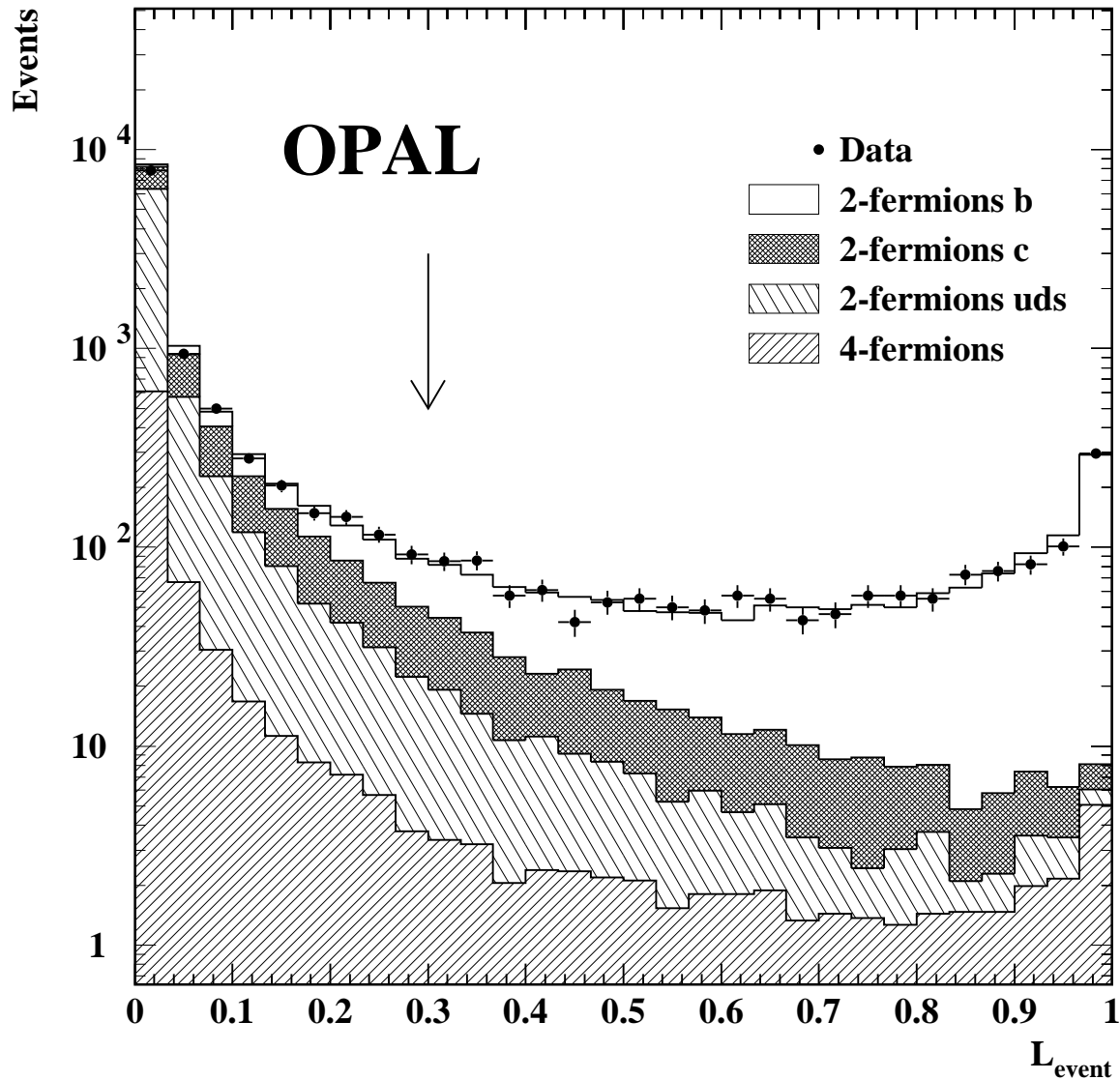}
}

\caption{ 
 ${\cal L}_{\rm event}$ distributions for selected non-radiative
  hadronic events. The points represent the data (182--209 GeV) and the
  histogram represents the Monte Carlo prediction, normalized
  according to the integrated luminosity of the data, with the open
  area representing the expected $\bbbar$  content and the
  hatched areas representing the contributions from
  $\ccbar$ light-quark pair and four-fermion events. 
  The arrow indicates the cut used.
}
\label{fig:lb} 
\end{figure}

In each hemisphere, the output of the lepton tag, the lifetime ANN and the 
kinematic ANN are combined with an unbinned likelihood calculation \cite{bib:pr261}, and the likelihoods
${\cal L}_{\rm hemi1}^{\rm b}, {\cal L}_{\rm hemi1}^{\rm c}  , {\cal L}_{\rm hemi1}^{\rm uds} , {\cal L}_{\rm hemi2}^{\rm b}, {\cal L}_{\rm hemi2}^{\rm c} , 
{\cal L}_{\rm hemi2}^{\rm uds}$ are obtained. The likelihood 
${\cal L}_{\rm hemi1}^{\rm b}$ 
is the probability for the first hemisphere to contain a b quark. 
${\cal L}_{\rm hemi2}^{\rm b}$ is 
the same probability for the second hemisphere, and  ${\cal L}_{\rm hemi1(2)}^{\rm c(uds)}$ 
are the corresponding probabilities for c-quark (light quarks).
The two hemispheres' outputs are then combined into a single event b-tagging likelihood 
variable ${\cal L}_{\rm event}$, using
$$  {\cal L}_{\rm event}=\frac{r_{\rm b}~{\cal L}_{\rm hemi1}^{\rm b}{\cal L}_
{\rm hemi2}^{\rm b}}{r_{\rm b}~{\cal L}_{\rm hemi1}^{\rm b}{\cal L}_{\rm hemi2}^
{\rm b}+r_{\rm c}~{\cal L}_{\rm hemi1}^{\rm c}{\cal L}_{\rm hemi2}^{\rm c}+
r_{\rm uds}~{\cal L}_{\rm hemi1}^{\rm uds}{\cal L}_{\rm hemi2}^{\rm uds}}.$$ 
The normalization 
parameters are set to $r_{\rm b}=0.165, r_{\rm c}=0.253$ and $r_{\rm uds}=0.582$, which are the 
$\ZFITTER$ theoretical predictions for $\Rb, ~\Rc$ and $\Ruds$ at about 190 GeV. The same values
are used for all data sets.
The number of events, $N_{\rm tag}$, satisfying a b-tagging cut of 
${\cal L}_{\rm event} >0.3$ is determined;
this cut value minimizes the total uncertainty in the measurement of $\Rb$.
Typical 
efficiencies for the b-tagger in the hadronic event sample are $65\%, 6.3\%$ and $1.5\%$ 
for $\bbbar, 
\ccbar$ and light-quark pair events respectively; for the residual four-fermion events the tagging efficiency is about 
5\%. On average, 77\% of the tagged sample is expected to be $\bbbar$ events.
The distribution of ${\cal L}_{\rm event}$ for the 182--209 GeV data is shown in 
Figure \ref{fig:lb}, together with the expectation from the Monte Carlo.
Good agreement between the data and the Monte Carlo is observed.

\subsection{Background estimation}
\label{ff}
The residual four-fermion background is estimated from Monte Carlo
simulation and subtracted from the number of selected hadronic ($N_{\rm had}$) and
b-tagged ($N_{\rm tag}$) events. It is 6.5\% for the hadronic event sample and 2.5\%
for the b-tagged sample. The dominant contribution to
the four-fermion background originates from $\wpwm$ and $\PZz\PZz$ final states, with
only 5\% coming from non-$\wpwm$ and non-$\PZz\PZz$ events.

Backgrounds from two-photon collisions and $\epem\to\tau^+\tau^-$ events were 
found to be negligible. 
The selected event sample has a small contamination from radiative hadronic events 
where $\sqrt{s'/s}$
was overestimated. This contamination is estimated using Monte Carlo 
simulation and subtracted from $N_{\rm had}$
and $N_{\rm tag}$ when calculating $N_{\qqbar}$ and $N_{\bbbar}$. Since both the 
b-tagged and the hadronic samples include such contamination, the effect on the 
measured value of $\Rb$ is small. 
About 3\% of the selected hadronic events and 4\% of the 
b-tagged events have a true value of $\sqrt{s'}$ smaller than
$0.85\sqrt{s}$.

The contributions from non-radiative $\ccbar$ and light-quark-pair events that
are b-tagged are taken into account by the two negative terms in the
numerator of equation 1. The secondary production of heavy quarks,
commonly referred to as gluon splitting, can also lead to primary
light-quark-pair events being selected by the b-tagging algorithm. 
This contribution is fully accounted for by the efficiencies in equation 1 if the
rates of this production 
in the Monte Carlo are correct. These rates were measured 
at LEP1 \cite{bib:gbb,bib:gcc,bib:gccLEP}. We extrapolate the measured values to the higher energies
using theoretical predictions given in \cite{bib:seymour,bib:pythia,bib:ariadne}. 
We obtain production rates per hadronic event, $\gccd=0.064\pm0.010$ and 
$\gbbd=0.0065\pm0.0015$
at $\sqrt{s}=200$ GeV where the uncertainties include the uncertainty on the measurements and the variation of the theoretical models 
used for the extrapolation to higher energy. The default Monte Carlo values are $\gccd=0.032$ and 
$\gbbd=0.0053$. We correct 
the value of \gcc\ in the Monte Carlo according to the extrapolated value. 
Since the Monte Carlo value of \gbb\ agrees with the data 
we do not correct the Monte Carlo for \gbb\ but take the 
difference between the Monte Carlo value and the maximal extrapolated rate 
($0.0053-0.0080$) as the corresponding systematic uncertainty.

\subsection {Subtraction of interference between initial- and 
final-state photon radiation }
\label{ifi}
The interference cross-section is defined as the difference
between the cross-section including interference between
initial- and final-state photon radiation and the
cross-section excluding interference. It is calculated
using $\ZFITTER$ with the Standard Model value of the forward backward asymmetry, for $\epem\to\bbbar$, $\epem\to\ccbar$ and 
$\epem\to\uubar,~ \ddbar,~ \ssbar$, in bins of $\cos\theta$ where
$\theta$ is the angle between the fermion direction and the
direction of the $\rm e^-$ beam. 
The interference cross-section
can be positive or negative, depending on the value of
$\theta$. The thrust axis gives a good estimate of the fermion direction, so we
assume that the interference correction estimated in bins of $\cos\theta$ can
be applied in bins of $\cos\theta_{\rm thr}$.
The fractions of b, c and uds events in the
hadronic and b-tagged samples are estimated from Monte Carlo samples generated excluding
interference. The interference cross-section for each flavour is then
weighted according to the estimated flavour composition, and the resulting numbers 
of events
are subtracted from the number of b-tagged events and from the number of \qqbar\ 
events to obtain $N_{\rm \bbbar}$ and $N_{\qqbar}$, which appear in equation 1. 
These corrections are typically 0.3\% for the hadronic sample and $-0.5\%$ for the b-tagged 
sample. The uncertainty on the above procedure is discussed in Section \ref{sec:ifisys} 
and given in Table~\ref{tab-sys}.
\subsection {Calculation of $\Rb$}
\label{results}

To avoid dependence on the Monte Carlo angular 
distribution, 
the analysis is performed in bins of $\vert\cos\theta_{\rm thr}\vert$.
In each $\vert\cos\theta_{\rm thr}\vert$ bin, 
the efficiencies and corrected numbers of $\qqbar$ and $\bbbar$ events are
 determined. The efficiencies, $\epsilon_{\rm b,c,uds}$, are defined as the number of 
MC events of the
specific flavour b-tagged in the $\vert\cos\theta_{\rm thr}\vert$ bin, divided by the 
number of events of that specific flavour that were generated in that bin.

\begin{figure}
\begin{center}
\includegraphics[width=6.8in]{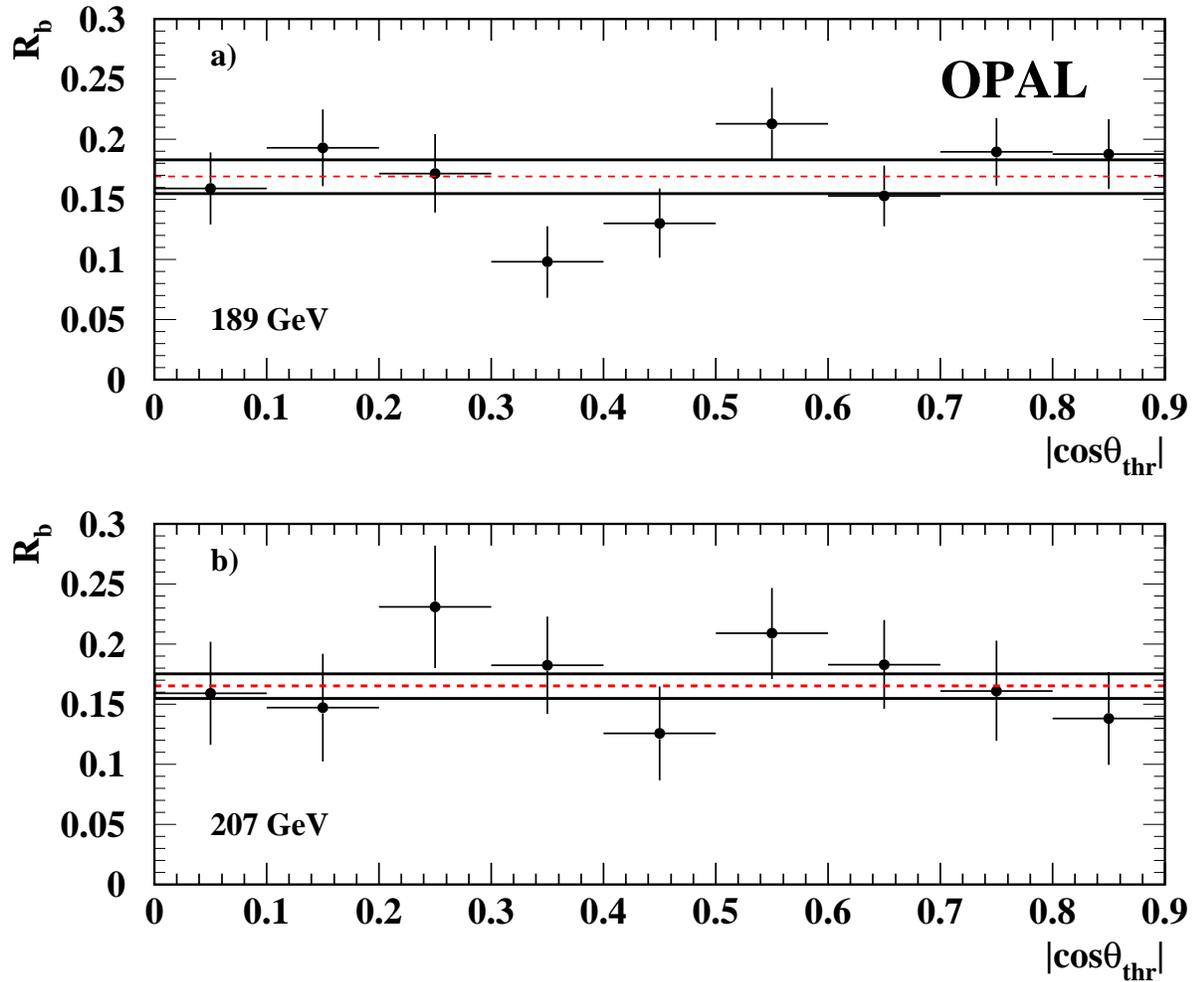}
   \caption{Measured values of $\Rb$ for: a) the 189 GeV sample; b) the 207 GeV sample with statistical uncertainties 
only. The dashed line is the average value and the solid lines show the 
the band corresponding 
to the one standard deviation uncertainty.
}
   \label{fig:Rb207}
\end{center}
\end{figure}

$\Rb$ is then calculated in each bin in the range  
$\vert\cos\theta_{\rm thr}\vert<0.9$.
 Figure \ref{fig:Rb207} shows the value of 
$\Rb$ versus $\vert\cos\theta_{\rm thr}\vert$ for the 189 GeV and 207 GeV samples.
The final value of $\Rb$ is determined by taking the average over all bins.
 The total numbers of selected hadronic and tagged events, the values of the variables in equation 1, as 
well as the resulting $\Rb$ value with its statistical uncertainty, 
are summarized in Table~\ref{tab:results}. 

\begin{table}[t]
 \begin{center} \begin{tabular}{c|cccccccc|c} \toprule
Data set & 
$ N_{\rm had} $ & 
$ N_{\rm tag} $ &
$ N_{\qqbar} $ &
$ N_{\bbbar} $  & 
$ \epsilon_{\rm b} $ & 
$ \epsilon_{\rm c} $ & 
$ \epsilon_{\rm uds}$ &
$ \Rc^{\rm SM}$ &
$ \Rb\pm{\rm stat}$  
\\
\bottomrule
183& 1172 & 185 & 1199.8 & 176.1 & 0.60 & 0.062 & 0.012 & 0.2532&$0.207\pm0.018$\\
189& 3496 & 447 & 3561.1 & 422.7 & 0.58 & 0.059 & 0.012 & 0.2551&$0.165\pm0.010$\\
192& \enspace573  & \enspace80  & \enspace584.7  & \enspace76.4  & 0.58 & 0.064 & 0.014 & 0.2560&$0.174\pm0.025$\\
196& 1343 & 182 & 1370.6 & 172.7 & 0.56 & 0.060 & 0.014 & 0.2570&$0.181\pm0.017$\\
200& 1312 & 166 & 1333.9 & 157.3 & 0.58 & 0.061 & 0.014 & 0.2581&$0.164\pm0.016$\\
202& \enspace591  & \enspace72  & \enspace604.6  & \enspace68.3  & 0.56 & 0.062 & 0.014 & 0.2585&$0.154\pm0.024$\\
205& 1124 & 136 & 1140.6 & 128.0 & 0.55 & 0.061 & 0.014 & 0.2592&$0.158\pm0.018$\\
207& 2046 & 251 & 2078.5 & 237.5 & 0.54 & 0.057 & 0.014 & 0.2597&$0.169\pm0.014$\\

\bottomrule

 \end{tabular}

 \caption {Number of selected ($N_{\rm had},~N_{\rm tag}$) and corrected ($N_{\rm \qqbar},~N_{\rm \bbbar}$) events; 
average (over all $\vert\cos\theta_{\rm thr}\vert<0.9$ bins) selection (including 
b-tagging) efficiencies; 
the Standard Model prediction for $\Rc$ 
 and the resulting value of $\Rb$ (with statistical uncertainties only) 
for the eight energies. Note that $\Rb$ is calculated from the values obtained in
bins of $\cos\theta_{\rm thr}$ and is not a direct calculation using the numbers in the table.}
\label{tab:results}
 \end{center}
\end{table}
The resulting $\Rb$ values are quoted for the case where 
$\Rc$ is set to its Standard Model value, $R_{\rm c}^{\rm SM}$, given in 
Table \ref{tab:results}.
The dependence of the result on the assumed value of $\Rc$ 
can be parametrized as:
$$  \Delta \Rb = b \; \left( \Rc - R_{\rm c}^{\rm SM} \right)
\,.$$
The parameter $b$, derived from equation 1, is determined separately for each
centre-of-mass energy and is given in Table~4. 

\section {Systematic uncertainties}
\label{sys}

The systematic uncertainties considered for these measurements are
described in the following paragraphs.
A breakdown of the systematic uncertainty at 207 GeV is
           given in Table~\ref{tab-sys}, and the total systematic error is given
           for all energies in Table 4.
\begin{table}[htb]
\begin{center}
\begin{tabular}{|l||l|}
 \hline
   Uncertainty source   & $\Delta R_{\rm b}/R_{\rm b}$ (\%) \\
\hline\hline
~~~b fragmentation & ~~~0.6 \\
~~~b decay multiplicity &~~~0.6 \\
~~~b hadron composition &~~~0.3 \\
~~~b lifetime           &~~~0.1 \\
~~~c fragmentation      &~~~0.8 \\
~~~c decay multiplicity &~~~0.1 \\
~~~c hadron composition &~~~0.3 \\
~~~c lifetime           &~~~0.1 \\
~~~semileptonic branching ratio &~~~0.1 \\
~~~semileptonic decay model &~~~0.1\\
Total b,c physics modelling    & 1.3 \\
$\rm K^0,~\Lambda$ rate     & 0.2 \\
Interference	                    & 0.3\\
Four-fermion background  &0.1 \\
\gbb 	            & 0.1 \\
\gcc	            & 0.1 \\ \hline
{\bf Total physics modelling}&{\bf 1.3}\\  \hline
Track reconstruction    & 3.2 \\ 
Lepton ID	            & 0.5  \\
Non-radiative event selection   & 0.1 \\
Monte Carlo statistics  & 0.6 \\\hline
{\bf Total detector effects}&{\bf 3.3}\\  
\hline\hline
{\bf Total systematic uncertainty } & {\bf3.5} \\ \hline 
\end{tabular}
\caption{\label{tab-sys}
Systematic uncertainty breakdown at $\sqrt{s}$ = 207 GeV. 
The uncertainties at other centre-of-mass
energies are similar. }
\end{center}
\end{table}


\subsection{Physics modelling}
\label{bcphmod.subsec} 
\subsubsection{Bottom and charm physics modelling}
Uncertainties in bottom and charm fragmentation and decay properties 
are discussed below. The variation of parameters is realized by reweighting
Monte Carlo events to the modified distribution.

\begin{list}{$\bullet$}{\setlength{\itemsep}{0ex}
                        \setlength{\parsep}{1ex}
                        \setlength{\topsep}{0ex}}
\item{\bf b fragmentation:} Although the mean scaled energy 
$2 \langle E_{\rm b} \rangle/\sqrt{s}$
of weakly decaying b hadrons is expected to change from LEP1 to LEP2 energies,
the free parameter $\epsilon^{\rm b}_P$ of the Peterson fragmentation function 
\cite{bib:peterson} used in the default MC for heavy flavour events is assumed not 
to vary with energy.
The value in \bbbar\ events is varied in the range  
$0.0030 < \epsilon^{\rm b}_P < 0.0048$,
which corresponds to a variation of the
mean scaled energy $2 \langle E_{\rm b} \rangle/\sqrt{s}$ 
of weakly decaying b hadrons in $\PZz$ decays in the range of
$2 \langle E_{\rm b} \rangle/\sqrt{s} = 0.702 \pm 0.008$ \cite{bib:frag-parameters}.
In addition, the heavy-quark fragmentation model is changed to that suggested by Collins and Spiller \cite{bib:CAS} 
and to that of Kartvelishvili \cite{bib:KAV}. The largest difference in all the above tests is from the Peterson parameter variation and is taken as 
the systematic uncertainty. 

\item {\bf b decay multiplicity:} The multiplicity of the charged decay products of 
hadrons containing a b quark is 
varied in the Monte Carlo simulation according to \cite{bib:frag-parameters}. 


\item {\bf b hadron composition:}
The tagging efficiency differs for the various b~hadron species.
The fractions of b~hadrons and their uncertainties 
have been taken from \cite{bib:PDG2002}.
The fractions $f({\rm B^0_s})$ and $f({\rm b_{baryon}})$
 are varied independently within ranges of the experimental measurements, 
and their variation
 is compensated by the fraction of $\rm B^0$ and $\rm B^\pm$ mesons.


\item {\bf b hadron lifetimes:} The lifetimes of the
different b hadrons are varied in the Monte Carlo by their
uncertainties according to \cite{bib:PDG2002}.


\item  {\bf c fragmentation:}  
Simulated $\rm c\overline{c}$ events have their
Peterson fragmentation parameter $\epsilon^{\rm c}_P$ varied in the range of
$0.022 < \epsilon^{\rm c}_P < 0.039$.
This corresponds to a variation of the
mean scaled energy $2 \langle E_{\rm c} \rangle/\sqrt{s}$ of weakly decaying 
c hadrons in $\PZz$ decays in the range of 
$2 \langle E_{\rm c} \rangle /\sqrt{s}= 0.484 \pm 0.010$ \cite{bib:frag-parameters}. 
\item  {\bf c decay multiplicity:} The average charged particle multiplicities
of $\rm D^+$, $\rm D^0$ and $\rm D^+_s$ decays are varied in the Monte Carlo 
within the ranges of the experimental measurements \cite{bib:mark3}.
\item  {\bf c hadron composition:}
The $\rm D^0$
fraction is written as 
$f{(\PDz)} = 1- f{(\PDp)} - f{({\PDsp})} - f{({\rm c}_{\rm baryon})} $.
The last three parameters are varied independently by their uncertainties
according to \cite{bib:lep_heavy}
to evaluate the uncertainty on the charm efficiency. 
\item   {\bf c hadron lifetimes:} Charmed hadron 
lifetimes are varied within their experimental uncertainties according to \cite{bib:PDG2002}.

\item {\bf semileptonic decay modelling and branching ratios:}
The semileptonic branching ratios of bottom- and charm-hadrons are varied within their 
experimental uncertainties \cite{bib:frag-parameters}. The uncertainty due to the semileptonic decay model used in the 
simulation is estimated as in \cite{bib:gcc}.

\end{list}

\subsubsection{Inclusive $\rm K^0$ and $\Lambda$ production rate}
The total production rates of $\rm K^0$ and $\Lambda$ in the Monte Carlo are varied by 4\% and 7\% respectively according to \cite{bib:double-tag}.
This variation contributes an uncertainty of 0.18\%.

\subsubsection{Initial-final state interference}
\label{sec:ifisys}
To asses the uncertainty on the correction for interference, we used the results of a study 
made in \cite{bib:twofermions} by replacing the \ZFITTER\ predictions with 
those of the KK2f program 
\cite{bib:kk2f}. We assign the difference between the two predictions 
as a systematic uncertainty.

\subsubsection{Four-fermion background}
 
The uncertainties on the measured  $\wpwm$ and  $\PZz\PZz$ cross-sections \cite{bib:ew}
and the uncertainty on the luminosity
               are found to have a negligible systematic effect on $\Rb$.
The background from W- and Z-pairs 
has the highest probability to be accepted in the tagged sample 
when one or both bosons 
decay into a final state containing a charm- or bottom-quark. 
The systematic uncertainty on the W- and Z-pair tagging efficiency is estimated by 
varying the tagging efficiency by 5\%. This variation introduces an uncertainty of 
0.15\%.
The effect of the detector resolution is also taken into account, 
as described in Section~\ref{detres.subsec}.

\subsubsection {Secondary heavy-quark production}
As discussed in Section \ref{ff}, the difference between the Monte Carlo rate for secondary \bbbar\ 
production and the highest expected value is taken as an uncertainty.
Typically, this process contributes a relative uncertainty of about 0.1\%.
%
The secondary \ccbar\ production rate is taken 
from \cite{bib:gcc,bib:gccLEP} and weighted according to \cite{bib:seymour} as 
described in Section \ref{ff}. The uncertainty on this estimation contributes a 
relative uncertainty of 0.13\%.

\subsubsection{Monte Carlo $\bbbar$ production}
The transition $N_{\rm had}\to N_{\rm \qqbar}$ has a small dependence on the Monte Carlo
value of $\Rb$. 
The default Monte Carlo value of $\Rb$ is varied within the measured uncertainty (Table 2). 
The effect on $\Rb$ is negligible.

\subsection{Detector effects}

\subsubsection{Track reconstruction}
\label{detres.subsec}

The detector performance in the Monte Carlo is varied as in \cite{bib:double-tag,bib:pr367}.
The effect of the detector resolution on the track 
parameters is estimated by degrading the resolution of 
all tracks in the Monte Carlo simulation.
This is done by applying a single multiplicative scale factor to the 
difference between the reconstructed and true track parameters.
A $5\%$ variation is applied independently
to the $r\phi$ and $rz$ track parameters. These variations together contribute 
an uncertainty of 1.8\%.
In addition, the efficiency for assigning measurement points
in the silicon microvertex detector to the tracks is varied by $1\%$
in $r\phi$ and $3\%$ in $rz$. Each of these variations contributes an uncertainty of 1.3\%.
The track reconstruction efficiency is varied by 1\%. This is done by randomly 
discarding 1\% of the tracks, and contributes an uncertainty of 1.8\%. 
The alignment of the silicon microvertex detector is changed in the Monte Carlo as in 
\cite{bib:double-tag}.
This variation gives an uncertainty of 0.4\%.
The systematic uncertainties resulting from the 
individual variations are summed in quadrature.

\begin{figure}[h]
\begin{center}
\includegraphics[width=6.5in]{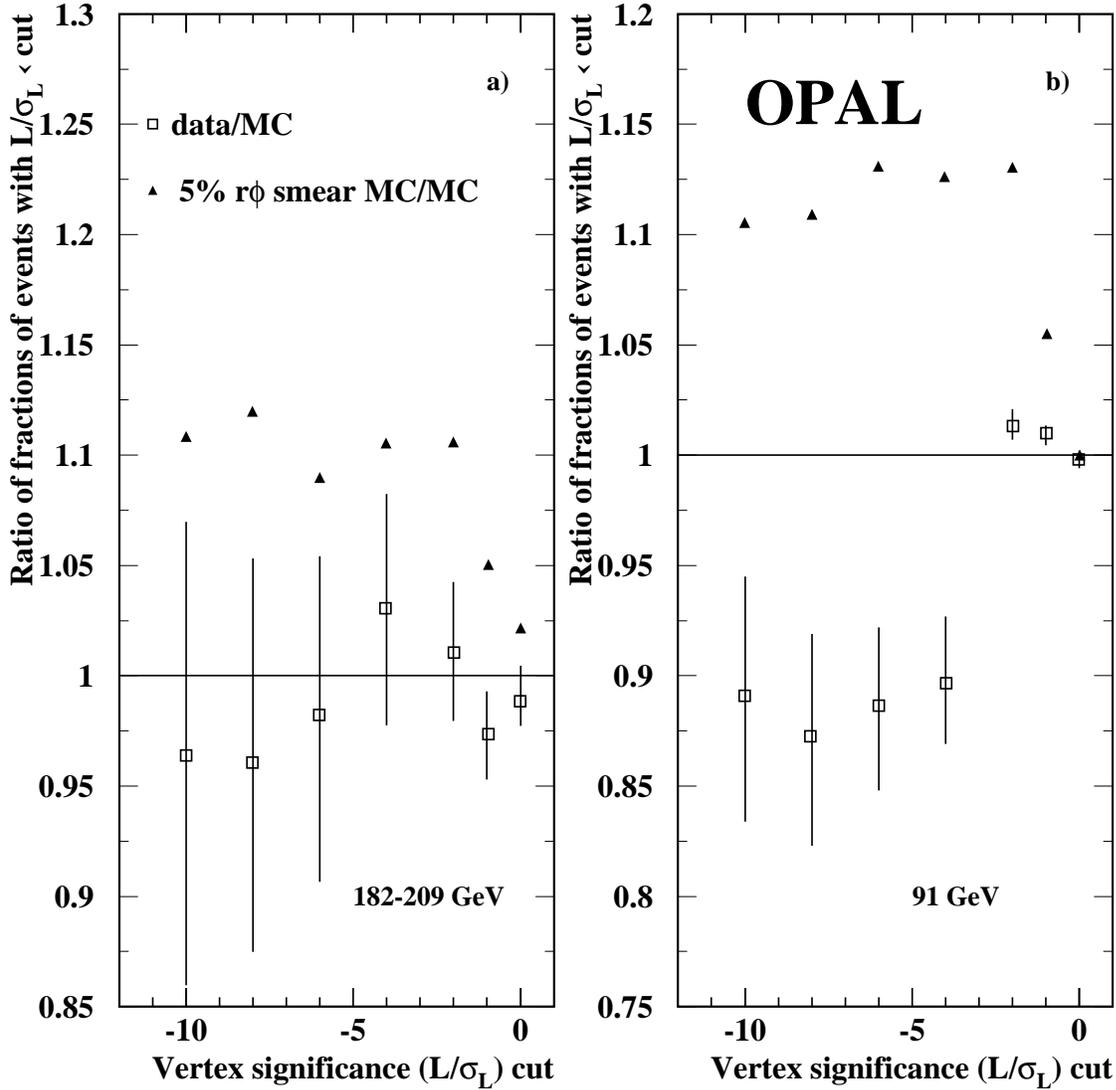}
\caption{Ratios of fractions of hemispheres with vertex significance
smaller than the cut value on the abscissa. The
data (open squares) and the Monte Carlo with 5\% deterioration of
the tracking parameters resolution in $r\phi$
(solid triangles) are normalised to the default Monte Carlo.
The uncertainties are a combination of the data and Monte Carlo
statistical uncertainties. Hemispheres are selected if the opposite
hemisphere in the event has a low value of $\cal{L}_{\rm hemi}$, giving a sample
75\% pure in uds events. The 182--209 GeV sample is shown in a) and the 91 GeV sample in b).}
   \label{fig:smear}
\end{center}
\end{figure}

Since the largest part of this uncertainty is due to the effect on 
$\epsilon_{\rm uds}$, we verify the validity of the 5\% variation in the $r\phi$ track parameters
by using events that are rejected by the b-tagger. We look at the 182--209 GeV sample 
as well as at data collected in the same years at  
the $\PZz$ peak for detector calibration.
Events  with a low value of $\cal{L}_{\rm hemi}$ in one hemisphere are selected, giving a
sample dominated by light-quark pairs. The value of the vertex
significance in the opposite hemisphere is then examined. Figure \ref{fig:smear} shows
the fraction of events with vertex
significance less than a negative cut value in the data and in the Monte Carlo with 5\%
variation in the $r\phi$ track parameters, normalised to the fraction in Monte Carlo.
This study suggests
that the difference between the data and the Monte Carlo is smaller than the variation
on the Monte Carlo and we conclude that the 5\% variation is adequate. 

\subsubsection{Lepton ID}
\label{leptonidsys.subsubsec}
The electron identification efficiency is varied by 15\% and the electron fake rate 
by 30\%. These values are 50\% larger than those used in \cite{bib:pr297} following a 
later comparison of OPAL's data and Monte Carlo. 
The muon identification efficiency is varied by 2\% and the fake rate by 14\%.
This source contributes typically an uncertainty of 0.45\% on $\Rb$
which is predominantly from the variation on the electron identification efficiency.

\subsubsection{Non-radiative event selection} 
We repeated the analysis using a different method for estimating $s'$, 
described in 
\cite{bib:pr158}. The resulting $\Rb$ was found to be different by
less than 0.1\%. This comparison is almost free of statistical uncertainties; thus we
take this difference as an uncertainty.

\subsubsection{Monte Carlo statistics}
The finite size of the Monte Carlo samples contributes between 0.6\% and 0.8\% 
uncertainty to the systematic error.

\subsection{Cross-checks}
\label{evselsys.subsubsec}
As a cross-check, the analysis is repeated on data collected in the same years at  
the $\PZz$ peak for detector calibration.   
A value of $\Rb(\sqrt{s}=m_{\PZz})=0.210\pm0.001({\rm stat}.) \pm 0.006 ({\rm syst}.)$ 
is obtained. The systematic uncertainty is determined as for the 
high-energy data samples. This result is consistent with the LEP1 
average value of $\Rb^0 = 0.21643 \pm 0.00073 $ \cite{bib:ew}.

\begin{figure}[h]
\begin{center}
\includegraphics[width=7.in]{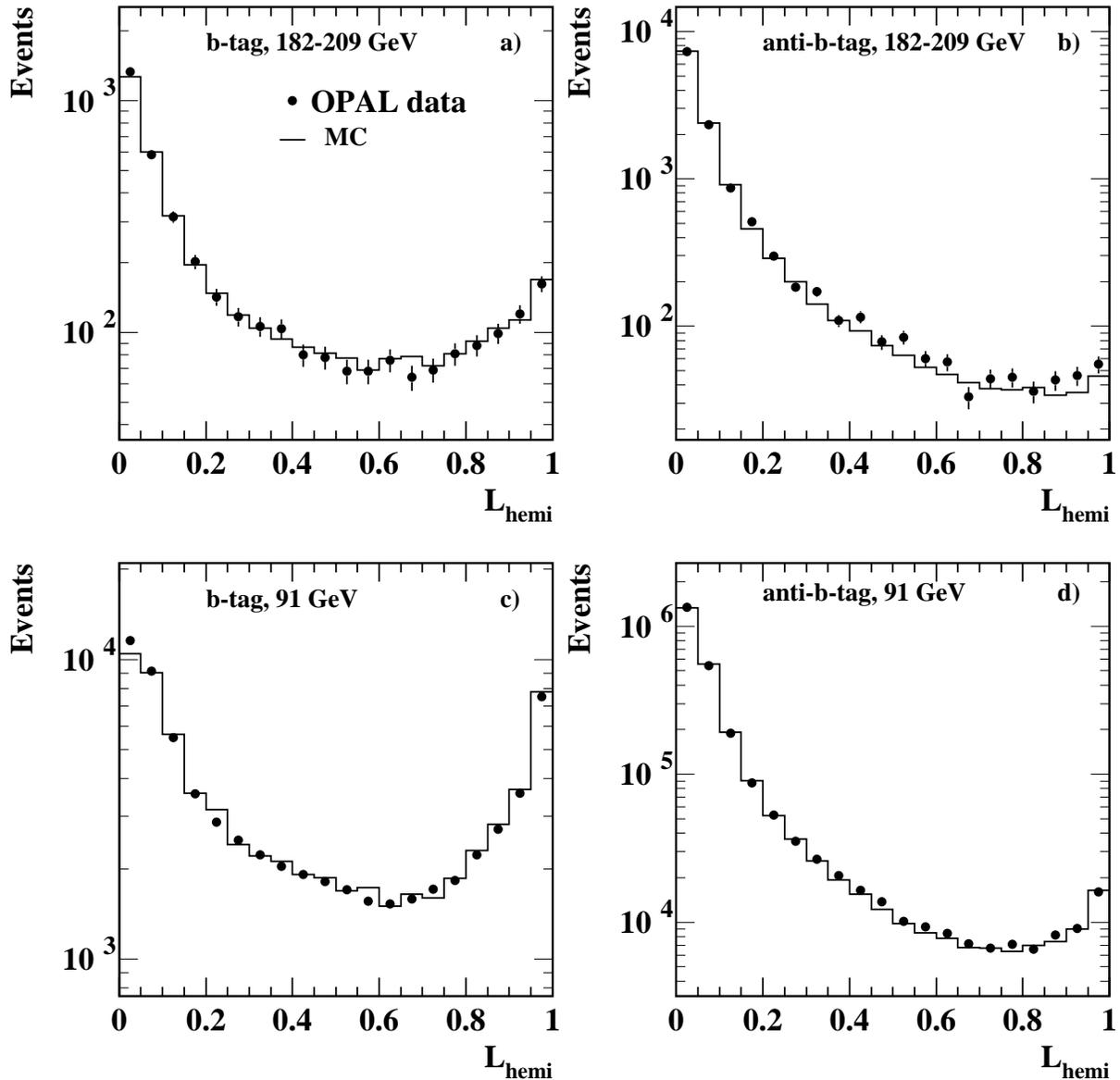}
   \caption{b-tagger output from one hemisphere for events selected by the opposite hemisphere. The top plots are for all data events in the range 182--209 GeV, and the bottom two are for the Z sample. a) and c) are for b-tagged opposite hemispheres and b) and d)
are for anti-b-tagged opposite hemispheres.}
   \label{fig:lbhemi}
\end{center}
\end{figure}

Another cross-check is made by comparing the likelihood of the hemisphere to contain 
a b-quark 
in Monte Carlo and data. This comparison is done for both the 182--209 GeV data and the 
calibration data at the $\PZz$ peak.
We select hemispheres where the b-tagger output in one hemisphere
indicates a high probability for the event to be a $\bbbar$ event, and look
at the opposite hemisphere.
The samples are 60\% and 88\% pure in $\bbbar$ events for the high energy and the Z peak 
data respectively. 
We also look at hemispheres in events that 
are rejected by the b-tagger in the opposite hemisphere, giving samples with only 6\% 
$\bbbar$ events. The results of this cross-check for the 182--209 GeV 
data and for the calibration data at the $\PZz$ peak are shown in 
Figure \ref{fig:lbhemi}. The small excess of data with respect to the Monte Carlo 
for ${\cal L}_{\rm hemi}>0.85$
in the anti-b-tagged plot of the 182--209 GeV sample (Figure \ref{fig:lbhemi} (b)) 
is covered by 
the detector resolution uncertainty. It is consistent with the result of the study 
shown in Figure \ref{fig:smear} (a). 
The difference between the Monte Carlo and the data in the b-tagged 91 GeV plot (Figure \ref{fig:lbhemi} (c)) is also covered by the detector resolution uncertainty.

\section{Results and conclusion}
\label{conclusion}
\begin{figure}[h]
\begin{center}
\includegraphics[width=6.8in]{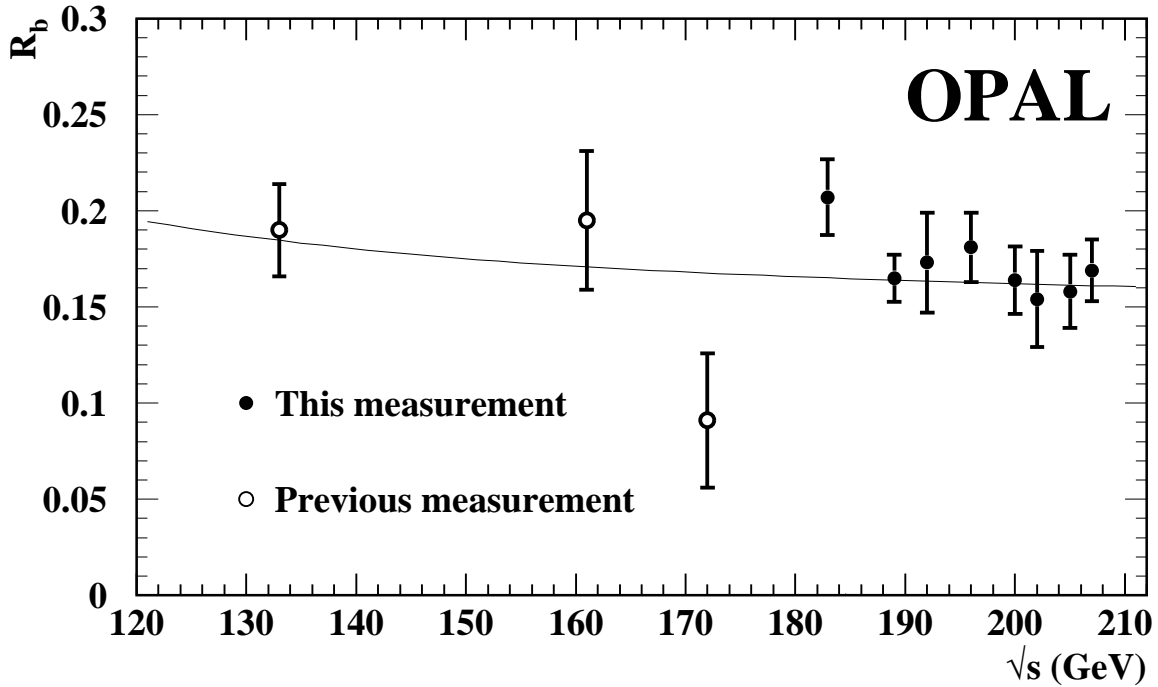}
   \caption{The measured $\Rb$ values (points with error bars) and the Standard Model 
prediction (solid line).}
   \label{fig:Rb}
\end{center}
\end{figure}

The $\epem\to\bbbar $ production rate per non-radiative $\epem\to\qqbar$ event has been
 measured 
using data collected by the OPAL detector at LEP at centre-of-mass energies 
between 182 GeV and 209 GeV.
The results are summarized in Table~4, together with earlier OPAL results from 130-172 GeV \cite{bib:pr297}. 
The measurements are shown in
                    Figure \ref{fig:Rb}, where they are compared with the
                    predictions of the Standard Model, calculated
                    using $\ZFITTER$.
Good agreement between the Standard Model and the measurements is observed. 
A comparison of the eight measurements reported here
to the Standard Model prediction gives: $\chi^2=5.0$, or 76\% probability of obtaining a larger 
difference than observed. Since the $\chi^2$ test only uses the absolute difference 
between the prediction and measurement, we also made another test assuming that the ratio 
of the measurement to the prediction is constant. We perform a $\chi^2$ fit for 
the ratio using only the statistical uncertainties and 
add the systematic uncertainties, which are assumed to be fully correlated, 
to the result. 
We obtain a ratio of $1.055\pm0.031~({\rm stat.})\pm0.037~({\rm sys.})$. 
Both tests suggest that these measurements are consistent with the Standard Model.
 
The measurement technique presented here has a larger selection 
efficiency and purity and thus yields a smaller statistical uncertainty and a
smaller dependence on $R_{\rm c}$ than the one used previously.   
The 183 GeV and 189 GeV results are in agreement with  \cite{bib:pr297}  
and supersede the previous measurements.

\begin{table}[h]
 \begin{center} \begin{tabular}{l|c|c} \toprule
$\sqrt{s}$ (GeV) & $\Rb\pm {\rm stat} \pm {\rm sys}$ & $b={\Delta\Rb\over(\Rc-\Rc^{\rm SM})}$ \\
\bottomrule

133.3 \cite{bib:pr297}	& $0.190 \pm 0.023 \pm 0.007 $&-0.12\\
161.3 \cite{bib:pr297}	& $0.195 \pm 0.035 \pm 0.007 $&-0.11\\
172.1 \cite{bib:pr297}	& $0.091 \pm 0.034 \pm 0.005 $&-0.11\\
{\bf  182.7}      	&{\bf $0.207 \pm 0.018 \pm0.007 $}&{\bf -0.085} \\
{\bf  188.6}	        &{\bf $0.165 \pm 0.010 \pm0.006 $}&{\bf -0.082} \\
{\bf  191.6}      	&{\bf $0.174 \pm 0.025 \pm0.006 $}&{\bf -0.088} \\
{\bf  195.5}	        &{\bf $0.181 \pm 0.017 \pm0.006 $}&{\bf -0.092} \\
{\bf  199.5}      	&{\bf $0.164 \pm 0.016 \pm0.006 $}&{\bf -0.083} \\
{\bf  201.6}      	&{\bf $0.154 \pm 0.024 \pm0.005 $}&{\bf -0.088} \\
{\bf  205.3}	        &{\bf $0.158 \pm 0.018 \pm0.006 $}&{\bf -0.088} \\
{\bf  206.8}      	&{\bf $0.169 \pm 0.014 \pm0.006 $}&{\bf -0.082} \\

\bottomrule
\end{tabular}
 \label{tab:conclusions}
\caption{$\Rb$ values with statistical (first) and systematic uncertainties. The last column
gives the dependence of the $\Rb$ value on $\Rc$. No variation of $\Rc$ is included
in the systematic uncertainty in the second column. The first three results
are from the previous OPAL measurement \cite{bib:pr297}.}
 \end{center}
\end{table}
\section{Acknowledgements}
\par
We particularly wish to thank the SL Division for the efficient operation
of the LEP accelerator at all energies
 and for their close cooperation with
our experimental group.  In addition to the support staff at our own
institutions we are pleased to acknowledge the  \\
Department of Energy, USA, \\
National Science Foundation, USA, \\
Particle Physics and Astronomy Research Council, UK, \\
Natural Sciences and Engineering Research Council, Canada, \\
Israel Science Foundation, administered by the Israel
Academy of Science and Humanities, \\
Benoziyo Center for High Energy Physics,\\
Japanese Ministry of Education, Culture, Sports, Science and
Technology (MEXT) and a grant under the MEXT International
Science Research Program,\\
Japanese Society for the Promotion of Science (JSPS),\\
German Israeli Bi-national Science Foundation (GIF), \\
Bundesministerium f\"ur Bildung und Forschung, Germany, \\
National Research Council of Canada, \\
Hungarian Foundation for Scientific Research, OTKA T-038240, 
and T-042864,\\
The NWO/NATO Fund for Scientific Research, the Netherlands.\\



\begin{thebibliography}{99}
\bibitem{bib:sm-rb}
G. Altarelli, T. Sj\"ostrand and F. Zwirner, {\it Physics at LEP2}
CERN-TH/{96-01}, Vol. 1, 1996.
\bibitem{bib:ew}
 {For a recent overview, see:\\
  The 4 LEP experiments: ALEPH, DELPHI, L3 and OPAL, the LEP Electroweak Working
  Group, and the SLD Heavy Flavour and Electroweak Groups,
  {\it A Combination of Preliminary Electroweak Measurements and Constraints 
       on the Standard Model, }
  CERN-EP/{2003-091}, December 2003 and references within.
  }

\bibitem {bib:twofermions}
OPAL Collaboration, G. Abbiendi \etal, Eur.~Phys.~J. {\bf C33} (2004) 173.

\bibitem {bib:double-tag}
OPAL Collaboration G. Abbiendi \etal,
Eur. Phys. J. {\bf C8} (1999) 217.

\bibitem{bib:pr297}
OPAL Collaboration, G. Abbiendi \etal, Eur. Phys. J. {\bf C16} (2000) 41.

\bibitem{bib:OPALdetectoratLEPtwo}
 {OPAL Collaboration, K. Ahmet et al., \NIM~{\bf A305} (1991) 275;\\
  O. Biebel et al., \NIM~{\bf A323} (1992) 169;\\
  M. Hauschild et al., \NIM~{\bf A314} (1992) 74;\\
  S. Anderson et al., \NIM~{\bf A403} (1998) 326;\\
G. Abbiendi et al.,
Eur. Phys. J. {\bf C14} (2000) 373.
 }
\bibitem{bib:kk2f}
  S.~Jadach, B.F.L.~Ward and Z.~W\c{a}s, Phys.~Lett {\bf B449} (1999) 97; \\
  S.~Jadach et al., \CPC\ {\bf 130} (2000) 260.

\bibitem{bib:grc4f}
  J.~Fujimoto et~al., \CPC\ {\bf 100} (1997) 128.

\bibitem{bib:koralw}
  S.~Jadach et al., \CPC\ {\bf 119} (1999) 272.
  
\bibitem{bib:pythia}
  T.~Sj\"ostrand et al., \CPC\ {\bf 135} (2001) 238.

\bibitem{bib:peterson}
C.~Peterson, D.~Schlatter, I.~Schmitt and P.~M.~Zerwas,
\PhysRev~{\bf D27} (1983) 105.

\bibitem{bib:frag-parameters}
   The LEP Collaborations, ALEPH, DELPHI, L3 and OPAL, Nucl. Instrum. Methods
   {\bf A378} (1996) 101.

Updated averages are described in ``Final input parameters for the LEP/SLD
heavy flavour analyses'', LEPHF/2001-01
(see {\tt http://www.cern.ch/LEPEWWG/heavy/} ).

\bibitem{bib:gopal}
OPAL Collaboration, J.~Allison \etal, \NIM~{\bf A317} (1992) 47.


\bibitem{bib:zfitter}
 D.~Bardin et~al., Comp. Phys. Commun. 133 (2001) 229; \\
   we use $\ZFITTER$ version 6.36 with default parameters, except 
  {\tt FINR}=0, {\tt BOXD}=2, {\tt CONV}=2 and {\tt INTF}=0 or =1, 
  and with the following input parameters: $\mPZ$=91.1852~GeV,
  $\mtop$=174.3~GeV, $\mHiggs$=115~GeV, $\Delta 
\alpha_{had}^{(5)}$=0.02761, 
  $\alphas(\mPZ)$=0.1185. These parameters are used for compatibility with \cite{bib:twofermions}.

\bibitem{bib:hsel}
OPAL Collaboration, G.~Alexander~\etal, \ZPhys {\bf C52} (1991) 175.

\bibitem{bib:172opal}
OPAL Collaboration, K. Ackerstaff \etal, Eur.~Phys.~J. {\bf C2} (1998) 213.

\bibitem  {bib:pr321}
OPAL Collaboration, G. Abbiendi \etal, Phys. Lett. {\bf B493} (2000) 249.

\bibitem {bib:pr367}
OPAL Collaboration, G. Abbiendi \etal, Eur.~Phys.~J. {\bf C26} (2003) 479.

\bibitem{muon}
OPAL Collaboration, R. Akers \etal, Z. Phys. {\bf C60} (1993) 199.

\bibitem{bib:pr146}
OPAL Collaboration, G. Alexander \etal, Z.~Phys.~{\bf C70} (1996) 357.


\bibitem{bib:Cpar}
G. Parisi, Phys. Lett. {\bf B74} (1978) 65; \\
J.F. Donoghue, F.E. Low and S.Y. Pi, Phys.~Rev.~{\bf D20} (1979) 2759.

\bibitem{bib:pr261}
OPAL Collaboration, G. Abbiendi {\it et~al.},
Eur. Phys. J. {\bf C7}  (1999) 407.

\bibitem{bib:gccLEP}
{
ALEPH Collaboration, R. Barate {\it et~al.},
Eur. Phys. J. {\bf C16}  (2000) 597;\\
L3 Collaboration, M. Acciarri {\it et~al.},
Phys. Lett. {\bf B476}  (2000) 243.
}

\bibitem{bib:gcc}
OPAL Collaboration, G. Abbiendi {\it et~al.},
Eur. Phys. J. {\bf C13}  (2000) 1.


\bibitem{bib:gbb}
{
ALEPH Collaboration, R. Barate {\it et~al.},
Phys. Lett. {\bf B434}  (1998) 437;\\
DELPHI Collaboration, P. Abreu {\it et~al.},
Phys. Lett. {\bf B405}  (1997) 202;\\
DELPHI Collaboration, P. Abreu {\it et~al.},
Phys. Lett. {\bf B462}  (1999) 425;\\
OPAL Collaboration, G. Abbiendi {\it et~al.},
Eur. Phys. J. {\bf C18}  (2001) 447;\\
SLD Collaboration, K. Abe {\it et~al.},
Phys. Lett. {\bf B507}  (2001) 61.
}

\bibitem{bib:seymour}
D.J. Miller, M.H. Seymour, Phys. Lett. {\bf B435} (1998) 213.



\bibitem{bib:ariadne}
  L.~L{\"o}nnblad, \CPC\ {\bf 71} (1992) 15.



\bibitem{bib:CAS} P. Collins and T. Spiller, J. Phys. {\bf G 11} (1985) 1289.

\bibitem{bib:KAV} V.G. Kartvelishvili, A.K. Likhoded and V.A. Petrov,
     Phys. Lett. {\bf B 78} (1978) 615.

\bibitem{bib:PDG2002}
Particle Data Group, K. Hagiwara \etal,
Phys. Rev. {\bf D66} (2002) 1.

\bibitem{bib:mark3}
MARK III Collaboration, D. Coffman \etal,
Phys. Lett. {\bf B263} (1991) 135.

\bibitem{bib:lep_heavy}
OPAL Collaboration, G. Abbiendi \etal,
Eur. Phys. J. {\bf C29} (2003) 463;\\
OPAL Collaboration, K. Ackerstaff \etal,
Eur. Phys. J. {\bf C7} (1999) 369.


\bibitem {bib:pr158}
OPAL Collaboration, G. Alexander \etal,
Z. Phys. {\bf C72} (1996) 191.


\end{thebibliography}
\end{document}